# Crystal Engineering and Ferroelectricity at the Nanoscale in Epitaxial 1D Manganese Oxide on Silicon

Andrés Gomez[1γ], José Manuel Vila-Fungueiriño[2,3γ], Claire Jolly[2], Ricardo Garcia-Bermejo[2], Judith Oró-Solé[1], Etienne Ferain[4], Narcís Mestres[1], César Magén[5], Jaume Gazquez[1], Juan Rodriguez-Carvajal[6], and Adrián Carretero-Genevrier[2]*.

1. Andrés Gomez, Dr. Judith Oró-Solé, Dr. Jaume Gazquez, Dr. Narcis Mestres.
Institut de Ciència de Materials de Barcelona ICMAB, Consejo Superior de Investigaciones Científicas CSIC, Campus UAB 08193 Bellaterra, Catalonia, Spain
2. Dr. Adrian Carretero-Genevrier, Claire Jolly, Dr. Ricardo Bermejo, Dr. José Manuel Vila-Fungueiriño.
Institut d'Electronique et des Systemes (IES), CNRS, Université de Montpellier, 860 Rue de Saint Priest 34095 Montpellier, France
3. Dr. José Manuel Vila-Fungueiriño
Centro Singular de Investigación en Química Biolóxica e Materiais Moleculares (CiQUS). C/ Jenaro de la Fuente s/n Campus Vida. Universidade de Santiago de Compostela. 15782, Santiago de Compostela, Galicia, Spain
4. Dr. Etienne Ferain,
it4ip S.A., avenue J.-E. Lenoir, 1, 1348 Louvain-la-Neuve, Belgium.
5. Dr. César Magén.
Instituto de Ciencia de Materiales de Aragón (ICMA), Universidad de Zaragoza – CSIC, Departamento de Física de la Materia Condensada, Universidad de Zaragoza, 50009 Zaragoza, Spain.
Laboratorio de Microcopías Avanzadas (LMA), Instituto de Nanociencia de Aragón (INA), Universidad de Zaragoza, 50018 Zaragoza, Spain.
6. Dr. Juan Rodriguez-Carvajal.
Institut Laue-Langevin, 71 Avenue des Martyrs, CS 20156, 38042, Grenoble, Cedex 9, France

γ These authors contributed equally to this work
E-mail: carretero@ies.univ-montp2.fr,

**Abstract**

Ferroelectric oxides have attracted much attention due to their wide range of applications, especially in electronic devices such as nonvolatile memories and tunnel junctions. As a result, the monolithic integration of these materials into silicon technology and its nanostructuration to develop alternative cost-effective processes are among the central points in current technology. In this work, we used a chemical route to obtain nanowire thin films of a novel $Sr_{1+\delta}Mn_8O_{16}$ (SMO) hollandite-type manganese oxide on silicon. Scanning transmission electron microscopy combined with crystallographic computing reveals a crystal structure comprising hollandite and pyrolusite units sharing the edges of their $MnO_6$ octahedra, resulting in three types of tunnels arranged along the c axis, where ordering of the Sr atoms produces a natural symmetry breaking. The novel structure gives rise to a ferroelectricity and piezoelectricity, as revealed by local Direct Piezoelectric Force Microscopy measurements, which confirmed the ferroelectric nature of SMO nanowire thin films at room temperature and showed a piezoelectric coefficient $d_{33}$ value of 22 ± 6 pC/N. Moreover, we proved that flexible vertical SMO nanowires can be harvested and converted into electric output energy through the piezoelectric effect, showing an excellent deformability and high interface recombination. This work indicates the possibility of engineering the integration of 1D manganese oxides on silicon, a step which precedes the production of microelectronic devices.

**Introduction.**

Complex oxides nanomaterials with ferroelectric, piezoelectric and multiferroic properties are pivotal to design devices such as sensors, micro energy harvesters or storage and magnetoelectric devices[1–4]. Moreover, to exploit their functionalities into future devices, the integration of these functional oxides into conventional semiconductor substrates in thin film form is essential[5–7]. A present-day example is the growth of a ferroelectric doped hafnium oxide directly on silicon[8], which could allow for polarization-driven memories. Other recent studies have reported a direct and bottom-up integration of a new family of epitaxial 1D single crystalline hollandite oxides on silicon substrates using chemical solution deposition (CSD) techniques[9]. The method relies on the thermal devitrification and crystallization of the silica native layer on the silicon surface into polycrystalline α-quartz, assisted by strontium alkaline earth cation (i.e. $Sr^{2+}$ or $Ba^{2+}$)[10,11]. The resulting α-quartz/Si interface act as a template to stabilize the crystallization and growth of 1D single crystalline hollandite nanowires [9].

Hollandites have the general composition of $A_xM_8O_{16}$, being A an alkaline earth cation and M a transition metal cation, and a crystal structure comprising double chains (zigzag chains) formed by edge-sharing $MO_6$ octahedra. Such double chains are corner-shared and form a $M_8O_{16}$ framework with tunnels [12]. Hollandites are under constant development, as shown by recent synthesis and design of novel structures with an extensive range of properties [13–19]. However, most hollandite oxides are only available as powder bulk or single crystal, limiting their practical application. The aforementioned bottom-up CSD synthesis on silicon overcomes these limitations[9] and offer the possibility to modify the chemical composition of hollandite nanowires and yield different physical properties[20–22] or even to stabilize novel crystallographic structures, such as the 1D $Sr_{1+\delta}Mn_8O_{16}$ (SMO) hollandite like structure[9]. In this work, we will show that using this chemical route one can grow large-scale 1D single crystalline and epitaxial SMO hollandite nanowire thin film directly on silicon. We will demonstrate that it is possible to construct a 3D structural model of the novel SMO hollandite consisting of a columnar mixture of hollandite and pyrolusite units by combining atomic-resolution scanning transmission electron microscopy images and computing programs existing in the *FullProf Suite (BondStr, Mol_tpcr)* and *CrysFML08 (Groups, Similar)*[23]. The 3D structural model reveals a spontaneous symmetry breaking due to the ordering of Sr atoms within the available tunnels that could explain the origin of the non-centrosymmetric character of the SMO, giving rise to ferroelectricity and piezoelectricity. For the later we have used Direct Piezoelectric Force Microscopy (DPFM), which is a powerful methodology recently developed by the authors that allows to directly measure the local piezoelectric effect in thin films using an atomic force microscope (AFM)[24,25]. DPFM measurements confirmed the ferroelectric nature at the microscale of this novel non-centrosymmetric tetragonal structure and showed a $d_{33}$ value of 22 ± 6 pC/N. In addition, we have also proved that this ferroelectric oxide can be used for energy harvesting applications. We present a first piezoelectric nanogenerator prototype based on vertical ultra-long SMO hollandites nanowires grown on a silicon substrate with an excellent deformability and high interface recombination and more importantly, they are lead free oxide materials composed of exclusively Sr and Mn cations which are the 12th and 9th most abundant metals in the Earth's crust.

## Results and discussion

### Synthesis, structural and chemical characterization of $Sr_{1+\delta}Mn_8O_{16}$ hollandite-like oxide nanowire films on silicon

SMO nanowire thin films on silicon were prepared by spin coating strontium rich solutions based on our previous development of ultra-long hollandite oxides nanowires[9]. Here, we applied a thermal treatment in air atmosphere at high temperatures (up to 750 °C) during 2 hours of the spin coated precursor's solutions containing strontium alkaline earth cations (see more details in the experimental section). At this temperature, strontium alkaline earth cations promoted the formation of a polycrystalline quartz layer at the silicon interface as previously observed[9,10] that finally stabilize the nucleation and crystalline growth of planar textured SMO nanowire flat films. Figure S1 shows a detailed tilted field emission gun scanning electron microscopy (FEG-SEM) image of the SMO nanowire film on silicon together with a cross-sectional high-angle annular dark field image (HAADF) scanning transmission electron microscopy (STEM) image of grown SMO nanowire film where the brighter rods correspond to the epitaxial SMO nanowires. The SMO nanowires of 1 µm length and 50 nm diameter are perfectly percolated with most of their longitudinal axis lying onto the polycrystalline α-quartz interlayer establishing a thin (50 nm thick) and flat film (roughness average of 4.4 nm) (see figure S1 b and c). A general schematic of the SMO nanowire film growth process together with FEG-SEM images acquired at intermediate growth steps are shown in Figure S2. At initial stages, i.e. at 700 °C (see Figure S2b), the SMO nanowires start to nucleate and grow, reaching an optimal crystallization after 2 hours at 750 °C where a continuous nanowire film is formed. Above this optimal growth temperature, i.e. at 800 °C, the SMO nanowire film is totally embedded within the quartz matrix. Temperature-resolved synchrotron grazing incidence 2D X-ray diffraction analysis allowed to observe the *in-situ* crystallization of epitaxial SMO nanowires thin films on silicon. The experiment was performed in the same conditions used at the chemical laboratory i.e. under air with a ramp temperature of 3 ºC min$^{-1}$ until 800 °C. Temperature-resolved synchrotron X-ray diffraction experiments (see Figure S3) confirmed that a homogeneous crystallization of SMO started at 700 °C without phase changes during the crystal growth. In a previous paper[9], we analyzed the electron diffraction patterns of the SMO crystalline nanowires that we attributed to a kind of SMO hollandite-type superstructure[26]. The electron diffraction patterns did not present ordered diffuse scattering or satellites indicative of a modulated structure. The calculated unit cell of the nanowires, called hereafter T-cell, corresponds to a tetragonal body centred cell with approximate lattice parameters $a_T \approx 25.2$ Å, $c_T \approx 5.7$ Å. At that moment, we had no electron diffraction pattern, or direct images, along the $c_T$-axis direction. The absence of this important structural information prevented any possibility to precisely reconstruct the atomic structure. Moreover, the amount, size and texture of SMO nanowire thin films is totally inadequate to resolve the complete atomic structure by the current techniques (X-ray diffraction, including synchrotron radiation, or neutron diffraction).

In the present work, we show the first atomic-resolution STEM images obtained with the electron beam along [001]$_{SMO}$, in which we are able to determine locally the chemical nature of the atomic columns. Combining high-resolution STEM imaging and computing programs existing in the *FullProf Suite*[23], we were able to construct the 3D structural model of this new structure. Figure 1a and 1b show a HAADF and a simultaneously acquired annular bright field (ABF) image along the [001] zone axis of a SMO nanowire, respectively. In the HAADF imaging mode the intensity of the image scales roughly with the square of the atomic number Z, which allows us to clearly distinguish between Sr and Mn atomic columns. On the other hand, the ABF imaging mode allows for visualizing the oxygen sublattice as well, and therefore complements the Z-contrast image. These two images unveil a structure consisting of a columnar combination of hollandite and pyrolusite units sharing the edges of their $MnO_6$ octahedra where the total number of Mn-atoms in the T-cell is 96 (see Figure 1c and Table S1 with the atomic positions of the model). While the pyrolusite tunnels remain empty, the hollandite and new tunnels generated in this structure are occupied by the Sr atoms. A comprehensive explanation of the methodology and reasoning employed for unraveling the structure of the SMO nanowires can be found in the supporting information.

We also analyzed the interfacial relationship between the polycrystalline quartz templating layer and the textured SMO nanowire film by atomically resolved STEM. Figure 1d shows a HAADF-STEM image of the interface between a single α-quartz crystal and a single SMO nanowire along the longitudinal nanowire axis, i.e. $[001]_{SMO}$ crystallographic direction. The fast Fourier transform (FFT) images reveal an in-plane epitaxial relationship between the SMO nanowire and quartz crystal given by (660) SMO [001] // (101) α-quartz [010] (see Figure 1e and f).). Notice that different spatial orientations of the quartz crystallites are possible therefore, the resulting SMO nanowires exhibit the same epitaxial relation but with diverse directions.

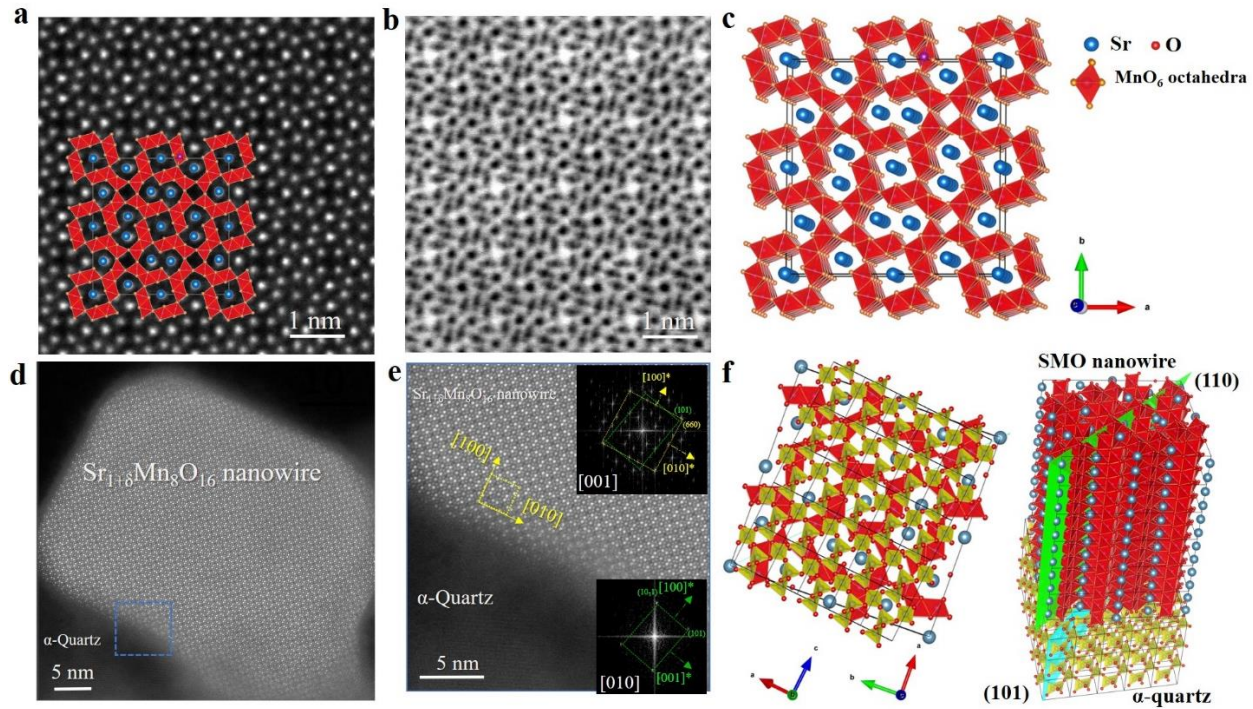

**Figure 1.** Structural characterization of epitaxial SMO nanowire film on silicon. (a and b) Atomically resolved HAADF and ABF-STEM images along the [001] zone axis of a single crystal SMO, respectively. The schematic atomic model is included in the (a) image. (c) Structural model proposed for a single SMO nanowire. Atomic positions are described in table S1 in the supporting information. (d) High resolution HAADF image of the SMO nanowire and quartz interlayer. (e) HAADF image of a single SMO nanowire with α-quartz interface. The insets show the FFT images that indicates the epitaxial relationship between both phases i.e. (660) SMO [001] // (101) α-quartz [010]. (f) Schematics of the atomic arrangement at the interface for the orientation relationship observed for SMO. In green the (110) plane of SMO nanowire parallel to the (101) plane of α-quartz highlighted in light blue.

Interestingly, Ph. Boullay *et al*., described a structure in the Ba-Mn-O system[27] that, in its [001] projected plane, is similar to what we observe for the pure Sr-Mn-O compound. The average structure in $Ba_{6-x}Mn_{24}O_{48}$ was described approximately in the *I*4/*m* space group with unit cell parameters $a_B$ =18.173(2) Å, $c_B$ =2.836(1) Å. However, their electron diffraction patterns indicate a more complex situation corresponding to a modulated composite structure with two different c-axes. Conversely, the diffraction patterns for the SMO structure do not present satellites or prominent diffuse scattering (see figure S4). The doubling of the c-axis with respect to the Ba-compound is well established in the electron diffraction patterns of the novel SMO structure (see figure S5). From the $I4/m$ tetragonal cell ($a_B$, $b_B$, $c_B$) proposed in[27], one can obtain the T-cell of SMO nanowires acquired with electron diffraction by applying the transformation: $a_T = a_B + b_B$, $b_T= -a_B + b_B$ and $c_T= 2c_B$. Therefore, one can deduce, using *Similar* software, that the non-centrosymmetric $I4$ subgroup is more appropriate to the SMO structure compared to the other possibility which is the centrosymmetric $I4/m$ (see atomic positions in table S1). This non-centrosymmetric character of the SMO structure can be attributed to an ordering of Sr atoms within the structural tunnels that we call tunnels C (see figure 2). Indeed, a full occupation by the strontium in tunnels C is not possible because there are many Sr-Sr distances equal to half c-axis ($\approx$ 2.87 Å) which is too short and electrostatically unfavourable. Moreover, the asymmetric unit in the non-centrosymmetric group $I4$ ($a_T$, $b_T$, $c_T$) contains 7 Sr, 12 Mn and 24 O independent atoms and only some Sr-atoms are in special positions (see Table S1). Considering a full occupation, the composition would correspond to the formula $Sr_5Mn_{12}O_{24}$. However, this composition is too high compared to the experimental relative cation composition obtained by means of EDX, this is $Sr_{0.11(2)}Mn_{0.64(9)}$, that in terms of hollandite formula becomes approximately $Sr_{1.4}Mn_8$ (see Figure S5). As a consequence, we have removed three Sr atom sites in $I4$ ($a_T$, $b_T$, $c_T$) that do not change the pattern of Sr observed in the high-resolution HAADF images. Therefore, the composition becomes $Sr_{1.8}Mn_8O_{16}$, which is similar to the composition measured by EDX if we allow a partial occupation in the hollandite tunnels. By removing these atoms, automatically we obtain a non-centrosymmetric structure because the presence of an inversion centre implies the presence of a mirror plane duplicating the Sr-atoms within the tunnels. Therefore, only the ordered Sr-atoms located in the tunnels C of the SMO structure break the centre of symmetry of the global structure. The rest of the Mn-O framework keeps the original inversion centre. In order to experimentally prove this hypothesis we have synthetized both, a (i) pure Ba hollandite oxide nanowire thin film that in our case exhibited the well known monoclinic structure ($Ba_{1.2}Mn_8O_{16}$, BMO) with cell parameters a = 10.052(1) Å, b = 2.8579(2) Å, c = 9.7627(10), and β = 89.96[28] (see Figure S6) and (ii) a Ba hollandite oxide nanowire thin film doped with 50% of Sr, $(BaSr)_{1+\delta}Mn_8O_{16}$ (BSMO), which disclosed the novel tetragonal structure above described in SMO nanowires system (see Figure S7). Remarkably, atomic-resolution electron energy loss spectroscopy (EELS) chemical mapping of the Mn, Ba, and Sr elements along the c axis of BSMO crystal structure showed unexpected cationic ordering within the novel tetragonal structure (see Figure 2 and Figure S7). Indeed, Ba atoms occupy all the hollandite tunnel (H) and prevent $Sr^{2+}$ ions to reside in these sites of the unit cell. Conversely, in the cationic positions corresponding to the C tunnels, one can observe that Ba and Sr share these sites therefore increasing the disorder in this position. This phenomenon is non-compatible with the *I*4 non-centrosymmetric structure due to the presence, on average, of a mirror plane within the tunnels that might fill partially the C tunnels with strong columnar disorder. As a result, the assembling of $Sr^{2+}$ and $Ba^{2+}$ cations within the tunneled structure is extremely important for the final functionality of the hollandite oxide, hindering, the ferroelectricity of BSMO nanowires as further described in the following section.

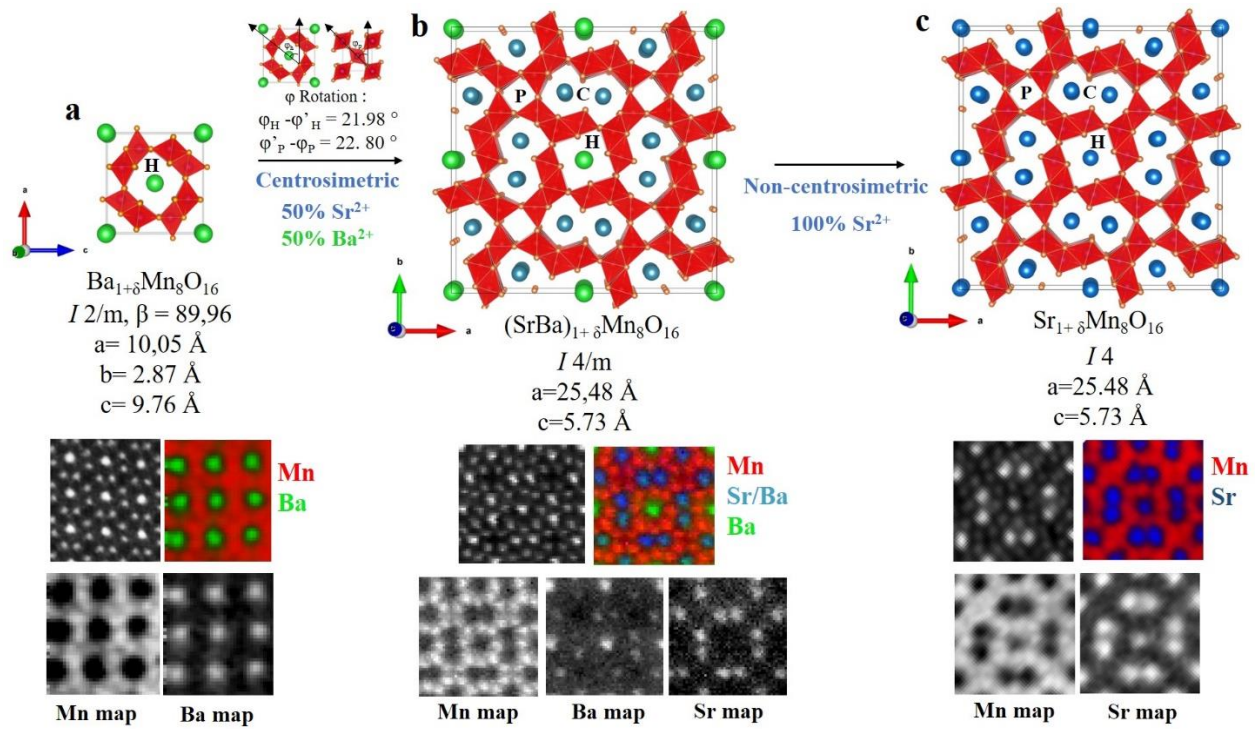

**Figure 2.** Crystallographic structures, composition and symmetry of: (a) BMO nanowire film (b) a BSMO nanowire film and (c) SMO nanowire film. Notice that in the novel SMO tetragonal structure, hollandites and pyrolusites blocks rotate 21.98 ° and 22.80 ° along φ angle respect to the well-known $Ba_{1+\delta}Mn_8O_{16}$ and pyrolusite ($MnO_2$) structures. The letter "H" in the structural models indicates that the atom belongs to the hollandite blocks or within the hollandite tunnels. The letter "P" indicates that the atom belongs to the pyrolusite blocks. The letter "C" indicates that the atoms are inside the new tunnels. The bottom panels of (a), (b) and (c) show the EELS elemental maps corresponding to the Mn L2,3, Ba M5,4 and Sr M4,5 edges along the [001] projection of the crystal structure of each individual nanowire unit cell.

**Ferroelectric characterization of SMO nanowire films on silicon**

We studied the ferroelectric properties at the micro and nanoscale of a 50 nm thick SMO nanowire film epitaxially stabilized on silicon with the DPFM technique [24]. For that purpose, we analyzed DPFM signals as a function of the scan direction on an antiparallel ferroelectric domain structure obtained by recording micro-range rectangular areas with a positive and negative sample bias of +20 and -20 VDC successively. Figure 3a describes the working principle of DPFM, where a metallic tip is used to scan the different ferroelectric domain structures induced in the SMO nanowire film; the blue color identifies a down domain polarization (noted as ''Pdw''), while the orange color recognizes an upwards polarization (noted as ''Pup''). The DPFM measurement was performed with the AFM working in contact mode. As the force is kept constant, the number of piezo-generated charges remains constant within the same ferro-electric domain, and it only changes when encounters and crosses a domain boundary. For instance, when the tip crosses the domain structures of figure 3a from left to right, a current is generated as the tip passes from the region of a negative charge accumulation (left side) to that of a positive charge accumulation (right side). When the tip scans back, the current changes its sign and flows in the opposite direction, as the tip goes from a positive charge accumulation (right side) to a negative charge accumulation region (left side). Such an imaging mechanism is unique because the only physical phenomenon that can produce such a contrast in an image is ferroelectricity. No other physical effect could induce a contrast reversal linearly dependent on the applied force and independent of the number of scans[24,25]. Figure 3c shows two images of a SMO nanowire film acquired scanning towards the left (upper panel) and towards the right (lower panel). Both present clear current peaks right at the antiparallel ferroelectric domain boundaries. Notice that both images are approximately mirror-like reflections of each other and that the current sign changes when scanning along different directions. The right to left scanning is referred to as DPFM-Signal Output (DPFM-So, upper panel) and the left to right scanning is referred to as DPFM-Signal Input (DPFM-Si, lower panel). Figure 3d shows two characteristic profiles, extracted from the DPFM images measured for a SMO nanowire film, that clearly demonstrates that the current is inverted when the scanning direction is reversed, therefore confirming the ferroelectric behavior of the SMO nanowire studied.

Indeed, the SMO nanowire film can be electrically switched because the polarization axis of the *I*4 tetragonal structure is parallel to the c axis, the longitudinal axis of nanowires, which generally lays flat on the silicon substrate. This allowed us to measure a similar ferroelectric domain structure with PFM mode at room temperature. Figure 3e shows vertical piezoresponse force microscopy (VPFM) phase and VPFM amplitude signals and proves the switching of the SMO nanowire film using a bias of ±20 VDV into an antiparallel natural domain structure configuration. Moreover, the profile extracted from the PFM images rules out any dielectric breakdown, see figure 3f. These results corroborate the previous DPFM measurements and thus the ferroelectric behavior of the SMO nanowire film.

In addition, we performed a comparative PFM study between the centrosymmetric BSMO, which displays a disorder of Ba and Sr atoms in the C tunnels and the non-centrosymmetric SMO in order to know if the particular cationic arrangement of the BSMO nanowires had any impact on the ferroelectric properties of these materials. Figure S8 and Figure S9 show the PFM study, which confirms the absence of ferroelectricity in the BSMO nanowire films. Therefore, we can conclude that the absence of ferroelectricity in the Ba doped SMO nanowire film is, most probably, due to the presence of Ba cations in the C tunnels, which increases the atomic disorder in this position and results in an average centrosymmetric structure. In this regard, a large number of studies have demonstrated the modification of several physical properties due to doping, such as the position of the foreign cation within hollandite tunnel structure[29].

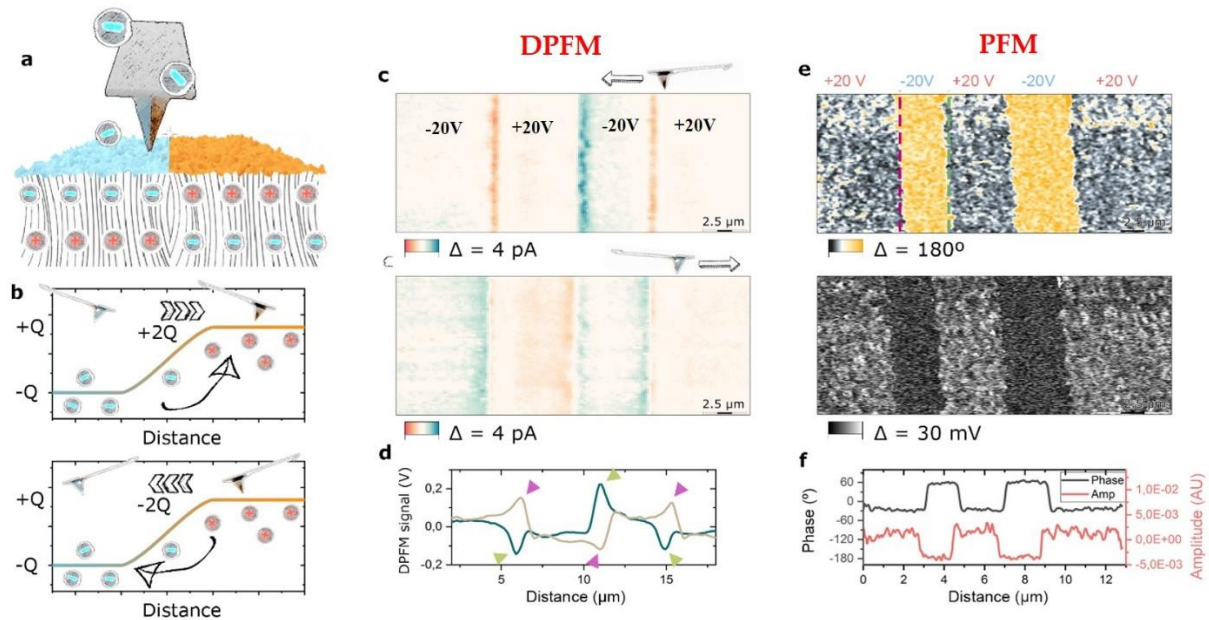

**Figure 3.** Ferroelectric characterization of SMO oxide nanowire films on silicon. (a) Scheme of the DPFM measurement of a SMO nanowire film, with an antiparallel up and downwards domain configuration. Upon application of a mechanical load, a negative charge is built up by the piezoelectric effect on the left side while a positive charge is built up at the right side (upper panel). The signal recorded, that is the current, is the derivative of the charge and it reverses its sign when the tip crosses different domains, depending upon the scan direction: the tip going from left to right and from right to left (lower panels, respectively), see (b). (c) DPFM images obtained for a SMO nanowire film with an antiparallel domain configuration. For a SMO nanowire film, the current sign is reversed as the scan direction changes, exactly as expected for a ferroelectric domain structure. (d) Random profiles extracted from a SMO nanowire film. (e) PFM analysis of the SMO ferroelectric antiparallel domain structure where VPFM phase and VPFM amplitude signals, upper and lower panel, respectively, confirm the switching of the nanowire film using a bias of ±20 VDV. (f) Random profiles extracted from the VPFM phase and VPFM amplitude signals of the SMO nanowire film.

**Piezoelectric characterization of SMO nanowire films on silicon**

Based on the fact that all ferroelectric materials are also piezoelectric[30], we studied the piezoelectric properties of SMO nanowire films using the DPFM technique. First, we maintained the AFM tip on the surface of the SMO nanowire thin film and simultaneously applied different force values, and then we measured the charges generated by the material. This allowed us to measure, with a high degree of precision, the piezoelectric constant $d_{33}$ of the SMO nanowire film, which is the charge generated per unit force. To quantify the $d_{33}$ value of SMO nanowire film, we directly integrate the peak current from the DPFM-Si image (see figure 4a). The values obtained were 7.3 fC and −8.4 fC, which divided by an applied force of 211 µN, yield a $d_{33}$ value of 22 ± 6 pCN$^{-1}$ (see figure 4a). The piezoelectric coefficient of SMO nanowire film was also measured using the conventional PFM technique, and it gave a $d_{33}$ value of 14 ± 2 pm/V (see Figure. S10). The two piezoelectric coefficients are of the same order, confirming the piezoelectric functionality of SMO nanowire films by two different techniques. Notice that α-quartz nanometric interfacial layer cannot contribute to the piezoelectric properties of SMO nanowire film since it is polycrystalline and a non-polar piezoelectric material with an extremely low $d_{33}$ value[31]. These are important results and suggest the possibility to engineer, by chemical routes, piezoelectric films directly on silicon with thickness ranging from 50 to 300 nm that may produce a high charge collection and more efficient MEMS energy harvester devices[32].

For a piezoelectric material, the recorded current should increase linearly with the applied load. Therefore, in the following experiments, we will show the response of the SMO nanowires film as a function of the applied force. Figure 4b shows the recorded DPFM-Si and DPFM-So images corresponding to a SMO nanowire sample under different applied loads, starting from a low loading force of 145 µN and then increased until reaching a maximum force of 269 µN. Notice that, when the force is increased, the current recorded by the amplifier increases as well, confirming the piezoelectric nature of the generated charge in SMO nanowire film. Figure 4c shows a hybrid piezo generated charge and topographic mapping of a SMO nanowire film on silicon, which confirms that the piezoelectric charges are generated homogeneously trough the nanowire film surface.

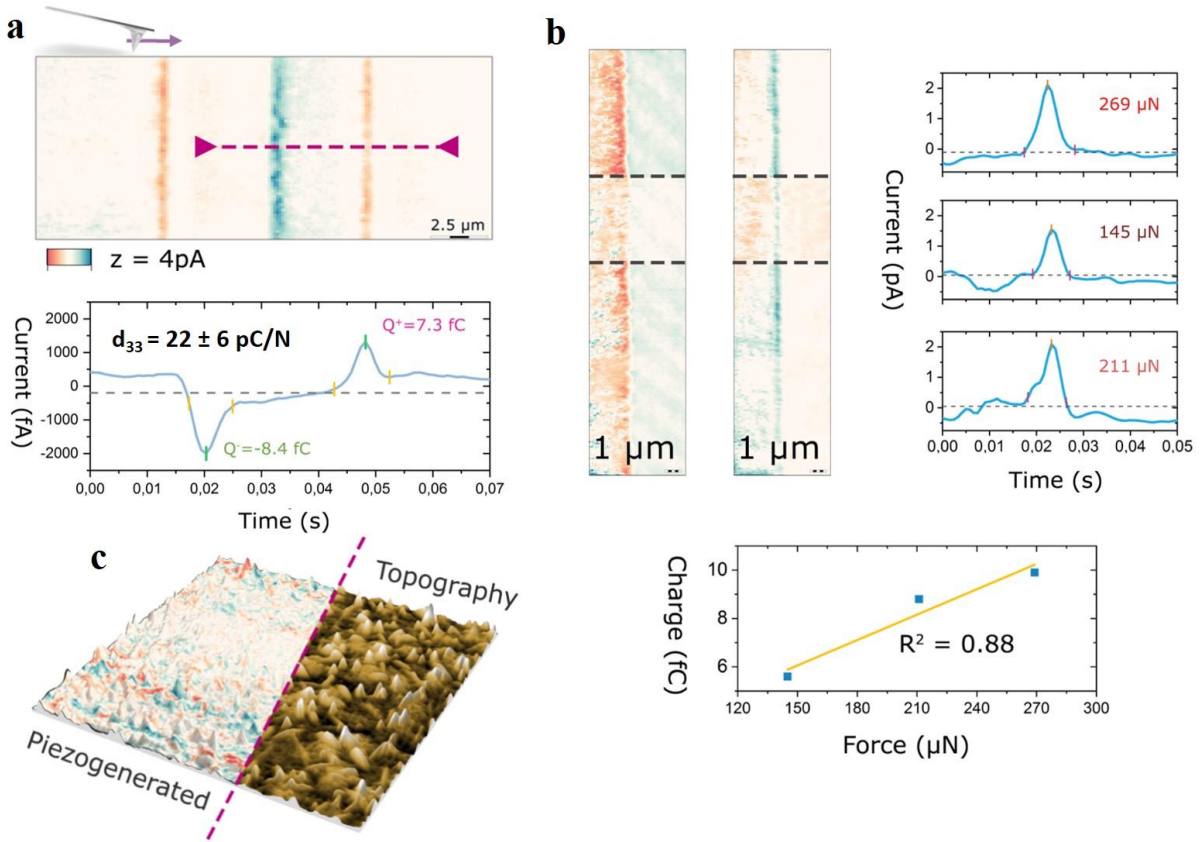

**Figure 4.** Piezoelectric characterization of SMO oxides nanowires film on silicon. (a) and (b) are recorded DPFM-Si and DPFM-So images, respectively, corresponding to a SMO nanowire film under different applied loads, starting from a low loading force of 145 μN till a maximum force of 269 μN. Lower panel in (a) shows the current vs. time profile along the dashed line in (a). Dashed lines in (b) separate regions in which different forces were applied. Right panel in (b) shows randomly selected profiles for the different applied forces. Lower panel in (b) shows the linear dependency of the force and the current recorded, confirming the piezoelectric nature of the generated charge in SMO nanowire film. (c) Hybrid piezo generated charge and topographic mapping of SMO nanowire film on silicon.

**Piezoelectric nanogenerators based on ultra-long SMO hollandite-like nanowires.**

Given the improved piezoelectric effect and excellent mechanical properties, one-dimensional (1D) piezoelectric nanostructures have been regarded as the next-generation piezoelectric material. In this light, 1D based nanogenerators can convert the mechanical energy into electricity by using piezoelectric 1D nanomaterials therefore exhibiting great potential in microscale power supply and sensor systems[33]. Accordingly, we tested the strain-induced piezoelectric polarization of flexible vertical ultra-long SMO nanowires by bending piezoelectric SMO nanowires and measuring the current obtained. We synthetized vertical 15 µm length SMO nanowires on highly doped silicon substrate using a 1D confined growth[9] (Figure S11). Figure 5a shows a cross-sectional FEG-SEM image of an ultra-long epitaxial SMO nanowire film, and Figure 5b depicts the nanogenerator based on the bending of SMO nanowires induced by an AFM tip. The power generation mechanism involves the bending and the stretching of the SMO nanowires along the ferroelectric axis under a tensile state induced by the AFM tip, see the sketch Figure 5c. A piezoelectric potential difference induced by the strain appears in the SMO nanowires when the AFM tip exerts a force, which drives the carriers through an external load and accumulates them at the interface between the metallic AFM tip and the contacted nanowires. When the force exerted by the AFM tip is released, also does the tensile strain, then the piezoelectric potential difference disappears and the accumulated charges move back from the interface. Consequently, repeatedly bending and unbending the vertical SMO nanowires with a metallic AFM tip results in the generation of alternating current peaks. Figure 5c shows the functioning of the SMO nanowire nanogenerator, which converts the mechanical energy into electricity. It constitutes the first piezoelectric nanogenerator prototype based on vertical and ultralong SMO hollandites' nanowires grown directly on a silicon substrate. These results constitute a promising and complementary avenue towards novel 1D electric nanogenerators, most of them based on ZnO nanowires which exhibit a similar piezoelectric coefficient $d_{33}$ of 14 pm/V[34]. Moreover, ferroelectric SMO hollandite nanowires possess excellent deformability and can be removed from the silicon substrate and aligned through a gentle scratch[35] on flexible Kapton substrate for being electrically contacted, as shown in Figure S12). This flexible manipulation and transfer of SMO nanowires makes this system a very promising material for applications in microelectronics.

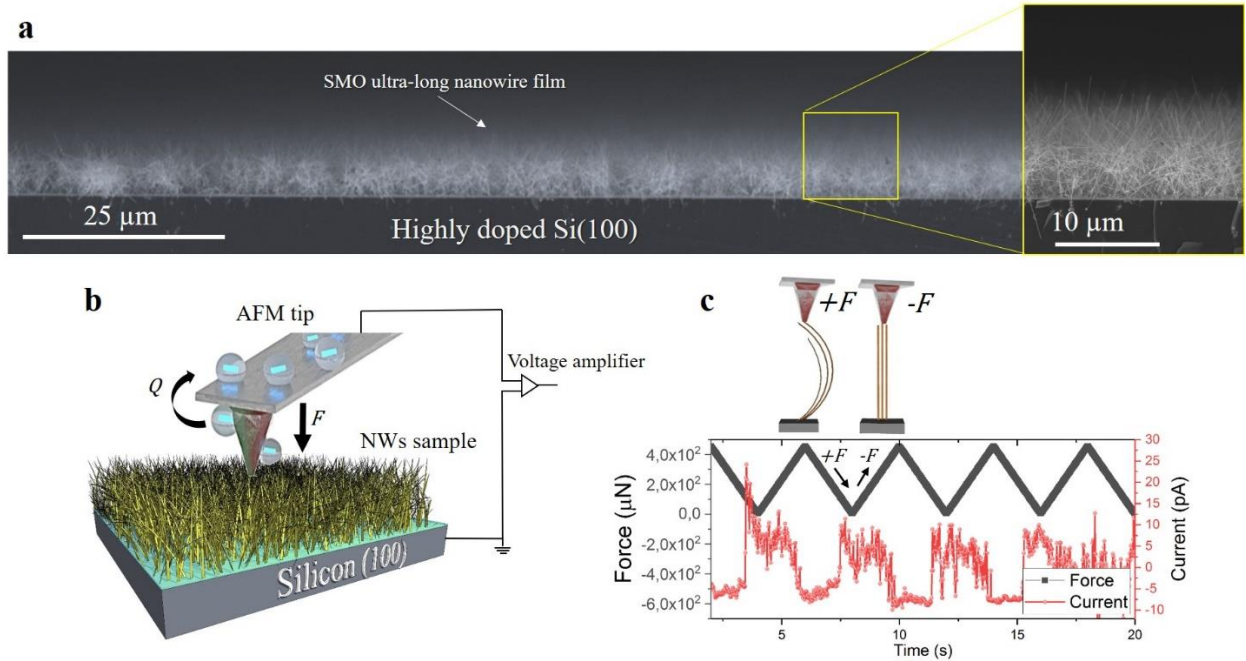

**Figure 5.** (a) cross-sectional FEG-SEM images of ultra-long epitaxial vertical SMO nanowire film on highly doped (100) silicon substrate. (b) sketch of the SMO nanogenerator based on nanowires bending induced by an AFM tip (c) The power generation mechanism involving the bending and unbending of the vertical SMO nanowire film with a metallic AFM tip, which results in the generation of the current peaks.

**Conclusions**

By using a cost-effective and scalable chemical method, we have integrated a room-temperature ferroelectric oxide in Si technology. Using a bottom-up chemical synthesis, we have been able to modify the chemical composition of hollandite nanowires directly grown on Si and to stabilize a new $Sr_{1+\delta}Mn_8O_{16}$ hollandite-like structure. The combination of scanning transmission electron microscopy and crystallographic computing revealed a novel structural model of $Sr_{1+\delta}Mn_8O_{16}$ nanowires comprising hollandite and pyrolusite units sharing the edges of their $MnO_6$ octahedra, resulting in three types of tunnels arranged along the c axis. The ordering of Sr atoms within the tunnels breaks the symmetry and gives rise to a ferroelectricity and piezoelectricity ($d_{33}$ value of 22 ± 6 pC/N), which have been locally examined using conventional and Direct-PFM techniques.

From a technological perspective, this novel hollandite oxide directly integrated on silicon opens the way for developing active devices engineered from led-free ferroelectric oxide materials. Indeed, we have used ultra-long and vertical $Sr_{1+\delta}Mn_8O_{16}$ nanowires as a nanogenerator prototype, which rendered an excellent deformability and high interface recombination. Besides, these ultra-long ferroelectric nanowires are easy to manipulate and transfer to flexible substrates making this system a very promising material for applications in microelectronics.

**Experimental Section**

**Ultra-long SMO nanowires synthesis.**

$Sr_{1+\delta}Mn_8O_{16}$, ultra-long nanowire films on silicon were synthesized using commercial nanoporous track etched polyimide (PI) templates (7µm thick and 200 nm of pore diameter) provided by it4ip s.a. (Beligium). This PI porous sheet is directly supported on the Si substrate and then impregned by capillarity forces with a mixed precursor solution containing 5 mL 1M Sr(Ac)$_2$ in acetic acid, 5 mL 1M of Mn(Ac)$_2$ in water and 10 mL of ethanol. Finally, a thermal treatment at 800 ºC for 5 h (ramp temperature 3 ºC min$^{-1}$) in air was applied directly in a tubular oven in order to obtain vertical epitaxial OMS nanowires on a silicon substrate.

**AFM characterization**

The measurements were performed using a Keysight 5500LS while a special ultra-low leakage amplifier is used in trans-impedance configuration (TIA) with part number ADA4530-1. The TIA amplifier is populated with a 10 GOhm feedback resistor, which give us a current noise level of 1:59 fA/Hz$^{0.5}$ while the leakage current in these conditions is maintained at less than 10 fA – data obtained from the amplifier datasheet of Analog Devices Inc. At the output of the current amplifier, a cascade voltage amplifier increases the overall gain of the system. The RC time constant of the amplifier is estimated at 6 ms, which equals a strain feedback capacitor of 0.1 pC. The two cascade amplifiers are calibrated with a known test resistor and a known bias applied, to see the calibration curves. In order to calculate the force used, we employed a standard method based on force vs. distance curves, performed in spectroscopy mode to acquire the tipde flection sensitivity. This value is used to convert the difference between the free deflection value and the set point value into force units, by using the cantilever spring constant. All measurements were performed in low humidity conditions, less than 8%, to avoid possible electrochemical effects.

**Structural and chemical characterization**

Hollandite like nanowires structure was investigated using a field emission gun scanning electron microscope (FEG-SEM), Hitachi's SU77. Scanning Transmission Electron Microscopy (STEM) studies were conducted on cross sectional TEM lamellas prepared by $Ga^+$ Focused Ion Beam (FIB) in a Thermo Fisher Helios Dual Beam. STEM analysis was carried out in a FEI Titan operated at 300 kV and equipped with a high-brightness field emission gun (X-FEG), a CETCOR aberration corrector for probe from CEOS GmbH and a Gatan Imaging Filter (GIF) Tridiem 866 ERS, and with a Nion UltraSTEM, operated at 100 kV and equipped with a Nion aberration corrector and a Gatan Enfina EEL spectrometer. In these microscopes, the aberration-corrected probe yields a routine spatial resolution below 1 Å, and the HAADF imaging allows recording incoherent Z-contrast images, in which the contrast of an atomic column is approximately proportional to the square of the average atomic number (Z). Reciprocal space reconstruction and determination of the different unit cells were investigated by using a JEOL 1210 transmission electron microscope operating at 120 KV, equipped with a side-entry 60/30º double tilt Gatan 646 analytical specimen holder and a link QX2000 XEDS element analysis system. X-ray diffraction measurements were carried out using a 6-circle diffractometer at Soleil Synchrotron equipped with a 2D detector Rayonix SX165. This detector is composed of a CCD camera, 165 mm diameter, 40 µm pixel size min. 4096x4096 pixels Pixel size: 40x40 µm$^2$ (min).

**Electrodes deposition and lithographic process**

Contact electrodes on aligned ultra-long SMO nanowires were performed by using a AZ2020 negative resist that was spin coated on Teflon substrate and patterned using EVG 620 UV lithography for the 50 nm of Cr and 120 nm of Au metal deposition. Lift off process was completed after 2 hours by using organic solvent photoresist remover PG at 80°C.

**Acknowledgements**

A.C-G, C.J, R.G-B and J.M.V-F. acknowledges the financial support from the European Research Council (ERC) under the European Union's Horizon 2020 research and innovation program (No.803004) and the French Agence Nationale de la Recherche (ANR), project Q-NOSS ANR ANR-16-CE09-0006-01. This project has received funding from the EU-H2020 research and innovation Program under grant agreement No 654360 having benefitted from the access provided by ICMAB-CSIC in Barcelona within the framework of the NFFA-Europe Transnational Access Activity. This project has received funding from the European's Union Horizon 2020 research and innovation programme under Grant No. 823717-ESTEEM3., the Spanish Ministry of Economy and Competitivity through Project MAT2017-82970-C2-2-R, and the Aragon Regional Government through Project No. E13_20R (with European Social Fund). We acknowledge SOLEIL for provision of synchrotron radiation facilities and we would like to thank Pierre Fertey for assistance in using beamline Cristal. J.G. also acknowledges the Ramon y Cajal program (RYC-2012-11709). The authors thank D. Montero for performing the FEGSEM images. The FEGSEM instrumentation was facilitated by the Institut des Matériaux de Paris Centre (IMPC FR2482) and was funded by Sorbonne Université, CNRS and by the C'Nano projects of the Région Ile-de-France. The authors thank Frederic Pichot for his expertise and advice during nanowire lithographic process. The STEM microscopy work was conducted in the Laboratorio de Microscopias Avanzadas (LMA) at Instituto de Nanociencia de Aragon (INA) at the University of Zaragoza as well as in the Center for Nanophase Materials Sciences at Oak Ridge National Laboratories (ORNL), which is a DOE Office of Science User Facility.